\documentclass{aastex6}
\usepackage{natbib}

\usepackage{color}
\shorttitle{Twist-removal in flare-associated sunspots}
\shortauthors{Burtseva et al.}

\begin{document}

\title{Search for a signature of twist-removal in the magnetic field of sunspots in relation with major flares}
\author{Olga Burtseva\altaffilmark{1}, Sanjay Gosain\altaffilmark{1}, and Alexei A. Pevtsov\altaffilmark{1,2}}
\altaffiltext{1}{National Solar Observatory, Boulder, CO 80303, USA}
\email{oburtseva@nso.edu}
\altaffiltext{2}{ReSoLVE Centre of Excellence, Space Climate research unit, 90014 University of Oulu, Finland}

\begin{abstract}
We investigate the restructuring of the magnetic field in sunspots associated with two flares: the X6.5
flare on 6 December 2006 and the X2.2 flare on 15 February 2011. The observed changes were evaluated with respect to the so-called twist-removal model, in which helicity (twist) is removed from the corona as
the result of an eruption. Since no vector magnetograms were available for the X6.5 flare, we applied
the azimuthal symmetry approach to line-of-sight magnetograms to reconstruct the pseudo-vector magnetic
field and investigate the changes in average twist and inclination of magnetic field in the sunspot
around the time of the flare. For the X2.2 flare, results from the full vector magnetograms were
compared with the pseudo-vector field data. For both flares, the data show changes consistent with
the twist-removal scenario. We also evaluate the validity of the azimuthal symmetry approach on simple
isolated round sunspots. In general, the derivations based on the azimuthal symmetry approach agree with
true-vector field data though we find that even for symmetric sunspots the distribution of the magnetic field may deviate from an axially symmetric distribution.
\end{abstract}

\keywords{Sun: flares - Sun: magnetic fields - Sun: photosphere}

\section{Introduction}
Flares are often seen as a tumultuous breakup of a previously stable configuration leading to the release of a significant amount of energy, material eruption, particle acceleration, and
restructuring of the magnetic field. This magnetic restructuring may be a key for understanding
the physics of processes taking place in the solar atmosphere in response to flares. \cite{Hudson00}
suggested a coronal implosion scenario, when during a flare or CME eruption, the coronal field
contracts toward lower atmospheric heights to reduce the magnetic energy stored prior to eruption,
resulting in the field becoming more horizontal at the photospheric level. Since then several
authors reported an increase of the horizontal field at the polarity inversion line (PIL) during flares
\citep[e.g.,][]{Gosain10,Wang10,Petrie12}, rapid change in inclination angle
\citep[e.g.,][]{Gosain10,Gosain12,Liu12}, and other signatures of contraction of magnetic fields
\citep[e.g.,][]{Li09,Wang12}. These observed changes support the early interpretation of eruptions in
the framework of the tether-cutting model \citep[e.g.,][]{Moore01}.

Rapid, ``permanent'' (or long--lasting) changes associated with the impulsive phase of flares were
also observed in longitudinal fields. For example, \cite{Sudol05} found permanent post-flare changes
in line-of-sight (LOS) magnetic fields measured with the Global Oscillation Network Group (GONG) instruments.
Most of the changes took place in sunspot penumbrae, and occurred in less than 10 minutes during the
impulsive phase of the flare. They also noted that the changes in magnetic field could be associated
with a rapid penumbral decay (when a portion of penumbra suddenly vanishes after the flare) reported
by other observers \citep[e.g.,][]{Liu05}. \cite{Sudol05} and \cite{Liu05} found roughly an equal
number of cases of longitudinal flux increase and decrease as the result of a flare, which seems to
be at odds with the coronal implosion scenario. If the magnetic field contracts back to lower
atmospheric heights, the field should always become more horizontal. \cite{Petrie10} further
analyzed flare-related changes in GONG LOS magnetograms, and found that an increase/decrease
in longitudinal fields does not correspond straightforwardly to a decrease/increase in field inclination 
relative to the local vertical direction. However, the distribution of longitudinal field
increases and decreases at different parts of active regions was found to be consistent with
Hudson's loop-collapse scenario. It may be noted that flare-related changes identified only in
vertical/LOS or horizontal components of the magnetic field may be a subject to an ambiguity, since
either a change in the field strength or inclination angle or both could be responsible for
the changes in vertical and/or horizontal components.

The implosion scenario gained significant support in published literature. However, it does not
explain all (global) changes in magnetic field observed during the flares and hence, there is a need
to explore other scenarios. For example, \cite{Petrie12} and \cite{Kazachenko15} studied changes in
the vector magnetic field in NOAA active region 11158 during the X2.2 flare, and while they reported the
increase in the horizontal field at the PIL, which is consistent with the loops' contraction, they
also found an abrupt decrease in magnetic-field twist. To explain the latter, one might consider an
alternative scenario based on magnetic helicity transport through the solar atmosphere. Many flare
models employ the concept of twisted magnetic flux structures. The basic assumption is that the
pre-flare configuration has excessive twist and (some of) this twist is removed from the corona as
the result of eruption. The evolution of twist in the flux tube after the eruption could be
understood in the framework of the \cite{Longcope00} model of the emergence of
a twisted flux tube through the photosphere into the corona. \cite{Longcope00} concluded that the expansion of the coronal portion of the tube redistributes the twist between subphotospheric and coronal parts, which in its turn creates an imbalance of torque at the photosphere-corona interface. To maintain
the balance of torque, the coronal portion of the flux tube should have its twist reduced. The
evolution of magnetic twist in emerging active regions observed by \cite{Pevtsov03} was found to be
in agreement with \cite{Longcope00} predictions. \cite{Pevtsov03} \citep[see also][]{Pevtsov08} made
a conjecture that the subphotospheric portion of flux ``tubes'' forming sunspots and active regions
may serve as a reservoir of twist (helicity) for repeated flare and CME eruptions. When twist is
removed from coronal fields (for example, via flares or CMEs), subphotospheric fields will re-supply
twist and establish a new equilibrium in torque in about one day. This transport of twist may
exhibit itself as rotational motion of the sunspot relative to its central point. The time for
establishing a new equilibrium is largely determined by the Alfv\'{e}n speed for the subphotospheric
portion of the flux tube. \cite{Pevtsov12} described several cases of flares associated with rotating sunspots. The speed of sunspot rotation was slowing down prior to flares, but after the
flares, it was increasing again. \cite{Pevtsov12} interpreted this as an indication of twist
(helicity) transport through the solar atmosphere, when the sunspot rotation was a response to
removal of helicity from the coronal portion of the flux system. The importance of twist in flares is
also supported by recent observations: \cite{Ravindra11}, and \cite{Inoue11} who conducted limited
case studies and found that magnetic twist increases about one day or longer before flare onset and
quickly decreases after the flare.

Unlike the tether-cutting (TC) model, a twist-removal (TR) model allows for a rapid restoration of
conditions leading to the flare and/or CME, and it can explain a sequential, multi-day flaring
``spree'' of an active region. Several other phenomena such as rotating sunspots
\citep[][]{Brown03} and higher flare productivity for active regions with a strong pattern of
kinetic helicity below them \citep[][]{Reinard10} seem to support the TR-model. Thus, for example, Pevtsov's (2012) study of rotating sunspots suggests that the amplitude of rotation increases after a major burst of flares in an active region, which suggests that rotation could be a reaction to the removal
of twist from active region magnetic fields. Sudden removal of twist from the
magnetic field will make the field more vertical, contrary to the coronal impulsion scenario. 

Both TC and TR scenarios may take place on the Sun. Sudden removal of some twist, by itself, would decrease the azimuthal component of the field in a cylindrical flux tube, making the field more vertical. Removal of twist, however, also decreases the free energy of the field, which should deflate the field as noted by \cite{Hudson00}. Deflating the field could amplify any remaining tilt, perhaps restoring part of the field's azimuthal component.

In this work, we investigate the magnetic restructuring associated with two flares, the X6.5 flare on 6 December 2006 and the X2.2 flare on 15 February 2011. We evaluate the observed changes in a search for change in the twist of the magnetic field, which could support the twist-removal model. Since true-vector data are not available, in our study of the X6.5 flare, we use only LOS magnetograms. For this event, we employ the approach of azimuthal symmetry to reconstruct three components of (pseudo-)vector of magnetic field, and we use this data to investigate the changes in twist and inclination of (pseudo-)vector magnetic fields as the results of flares. Although the application of this method is limited to flares associated with relatively symmetric sunspots, it can be used for events not observed by vector magnetographs, such as, for example, the Helioseismic and Magnetic Imager (HMI) on board the Solar Dynamics Observatory (SDO) \citep{Pesnell12} or Hinode/SOT (Solar Optical Telescope; \cite{Kosugi07}). For the X2.2 flare, we verify the reconstruction results from the azimuthal symmetry approach with the derivations based on full vector magnetograms from HMI. We also evaluate the validity of the azimuthal approach on simple, round, isolated sunspots. The rest of the article is organized as follows. In Section \ref{sec:derivation}, we formulate the principles of azimuthal (cylindrical) symmetry for reconstructing three components of the pseudo-vector field from LOS data. Section \ref{sec:x6flare} discusses changes in orientation of
the pseudo-vector field in relation with the X6.5 flare on 6 December 2006. Section \ref{sec:x2flare}
describes and compares the changes in magnetic field derived both from true and pseudo-vector data.
Section \ref{sec:round} analyzes the applicability of azimuthal symmetry to magnetically isolated
round sunspots, and Section \ref{sec:discussion} presents the critical discussion of the main findings
of this work.

\section{Derivation of the pseudo-vector magnetic field}
\label{sec:derivation}

Let us describe a vector magnetic field in a cylindrical coordinate system with the origin at the
center of the sunspot, azimuth angle, $\varphi$, and distance from the origin, $r$, by \{$B_z$, $B_R$,
$B_{\varphi}$\}, the vertical (up-down), radial (inward-outward), and tangential (clock-counter
clockwise) components, respectively. Reference (zero) azimuth is toward solar disk center and
azimuth increases in the clockwise direction. Then, the line-of-sight component, $B_{LOS}$, can be written
as \citep[see] [for more details] {Pevtsov90}:

\begin{eqnarray}
\label{eq1}
B_{LOS}(r, \varphi) = \cos\theta B_z(r, \varphi) + \sin\theta \cos\varphi B_R(r, \varphi) - \sin\theta \sin\varphi B_{\varphi}(r, \varphi)
\end{eqnarray}         
          
where $\theta$ is the heliocentric distance of the sunspot. Assuming cylindrical symmetry, i.e., $B_z$, $B_R$, and $B_{\varphi}$ are only functions of $r$, but not $\varphi$, the three field components can be computed as

\begin{eqnarray}
\label{eq2}
B_z(r) & = & \frac{1}{2\pi\cos\theta} \int_{0}^{2\pi} B_{los}(r,\varphi) d\varphi \nonumber\\
B_R(r) & = & \frac{1}{\pi\sin\theta} \int_{0}^{2\pi} B_{los}(r,\varphi) \cos\varphi d\varphi \\
B_{\varphi}(r) & = & \frac{-1}{\pi\sin\theta} \int_{0}^{2\pi} B_{los}(r,\varphi) \sin\varphi d\varphi \nonumber
\end{eqnarray}

\begin{figure}[h]
\begin{center}
\plotone{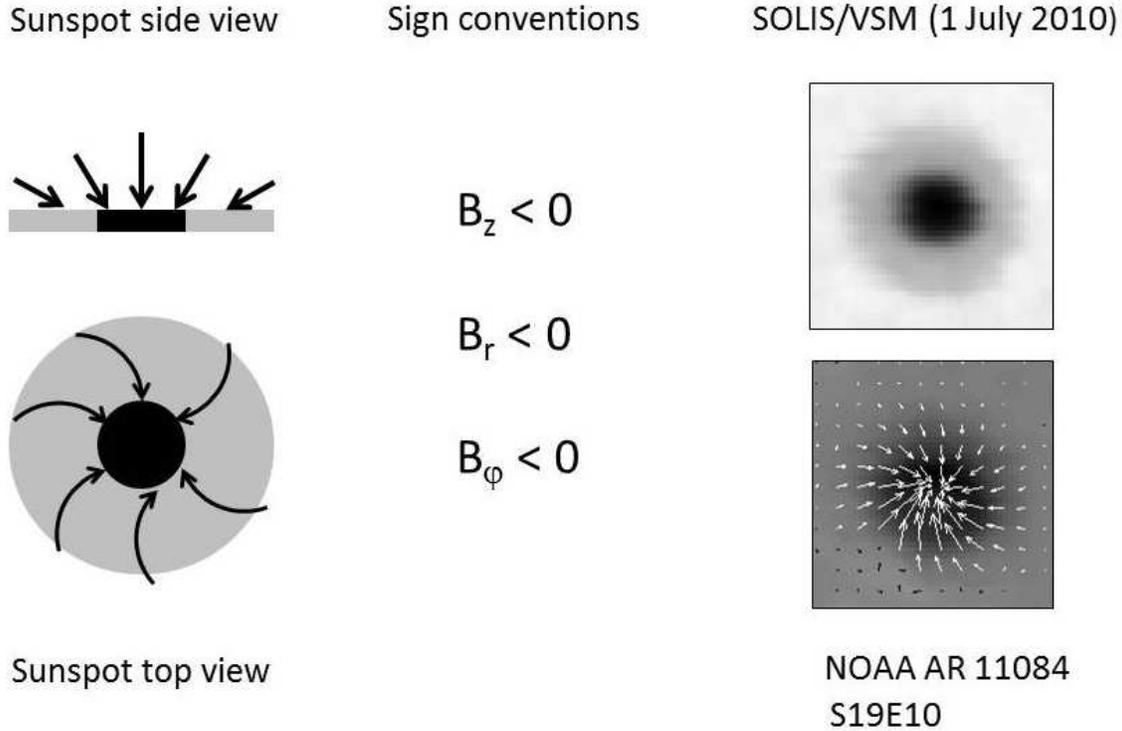}
\caption{Cartoon explaining the sign convention for pseudo-vector field components. SOLIS/VSM
continuum intensity at the top right and true-vector field observations at the bottom right with the
vertical field component, $B_z$, as the background, with the horizontal components indicated by arrows. Dark halftones correspond to the negative polarity field, and the length of the arrows is proportional to horizontal field strength.}
\label{fig1}
\end{center}
\end{figure}

The cartoon in Figure \ref{fig1} represents the  sign convention for the pseudo-vector field components:
negative $B_z$ and $B_R$ are in the downward and ``into the sunspot'' direction, respectively, and
negative $B_\varphi$ corresponds to the clockwise direction. Figure \ref{fig1} also shows an example of
distribution of the horizontal components of the magnetic field in round sunspot NOAA 11084 with the
same sign (as in the cartoon) of $B_z$, $B_R$, and $B_\varphi$.

The pseudo-vector calculations could be used to investigate twist reduction and loop collapse via
changes in the $B_z$, $B_R$, and $B_{\varphi}$ components, and field component
$B_h$=$\sqrt{B_R^2+B_{\varphi}^2}$. Twist reduction would imply a decrease in $B_{\varphi}$ and
possibly an increase in $B_z$. Loop collapse would give an increase of $B_h$ (and possibly a
decrease of $B_z$). The pseudo-vectors could help to discriminate between loop collapse and twist
reduction with a large number of pre-SDO (and, in particular, GONG) observations, which would be
important to this cycle with such a relatively small number of flares. Two of such major flares are
discussed in more detail in the following sections.

\section{Flare-related magnetic field changes: X6.5 flare on 6 December 2006}  
\label{sec:x6flare}

\begin{figure}[h]
\begin{center}
\plotone{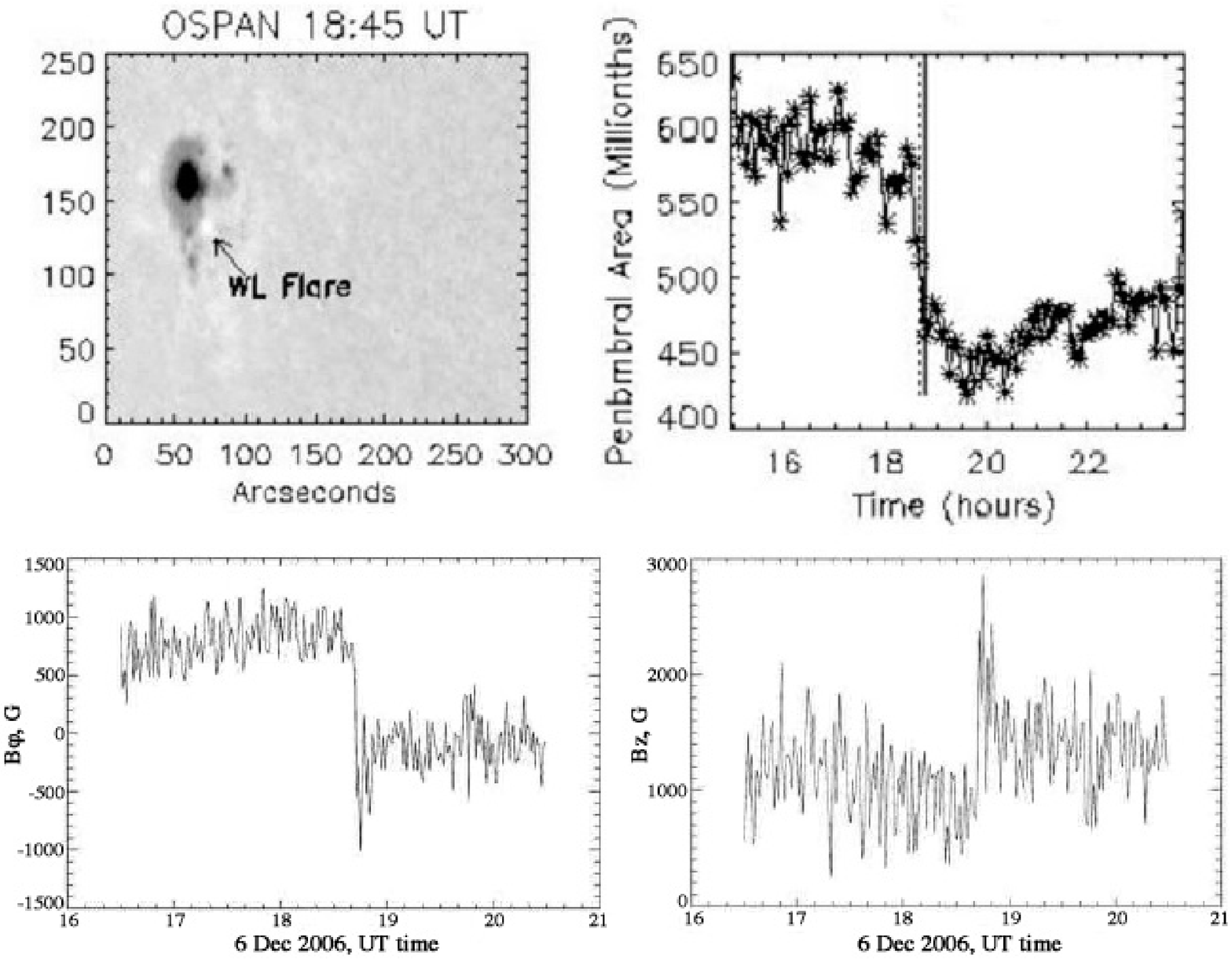}
\caption{White light image from OSPAN (upper-left) with a white light flare at about 18:45 UT and a
corresponding step-wise reduction in area of penumbra (upper-right). Two lower panels show time
evolution of vertical ($B_z$, lower-right) and azimuthal ($B_{\varphi}$, lower-left) components of
the pseudo-vector reconstructed under the assumption of cylindrical symmetry. Vertical lines in the upper-right panel mark start and peak times of this flare in white light intensity.}
\label{fig2}
\end{center}
\end{figure}

The X6.5 flare on 6 December 2006, which occurred in NOAA active region 10930, was a white-light flare,
associated with a Moreton wave and coronal mass ejection \citep{Bala10}. In the upper panel in Figure \ref{fig2}, the image from the Optical Solar Patrol Network (OSPAN, later known as the Improved Solar Observing Optical Network, ISOON \citep{Neidig98}) shows a white light flare at about 18:45 UT and a stepwise reduction in penumbral area (courtesy of \cite{Petrie12a}, see also \cite{Bala16}). The flare was observed by GONG instruments. GONG full-disk magnetograms are obtained with 1-minute cadence, 2.5-arcsec pixel size, and 3 G per pixel noise level. Previous studies reported significant stepwise changes in the GONG LOS magnetic field associated with the flare \citep{Petrie10, Burtseva13}. Analysis of the direction of the Lorentz force changes suggested the contraction of the field lines toward the neutral line resulting in a more horizontal magnetic field at the neutral line region \citep{Petrie10}. \cite{Burtseva13} found contraction of the field toward the neutral line in the form of flux cancellation events during the X6.5 flare, but no correlation between the flux cancellation and abrupt stepwise field changes associated with coronal field implosion. Using Hinode G-band white light and Ca {\rm II} H -- line broadband filtergrams,
\cite{Deng11} presented the rapid stepwise decay of the outer sunspot penumbra and enhanced, sheared Evershed flow near the flared neutral line, indicating weakening of horizontal field in the outer region and showing a more horizontal sheared magnetic field near the flared neutral line.

True-vector magnetic field observations at the time of this flare are not available, and thus, we
employ the cylindrical symmetry approach to reconstruct the vertical, radial, and tangential components
\{$B_z$, $B_R$, $B_{\varphi}$\} of the vector magnetic field following the approach presented in Section
\ref{sec:derivation}. Time evolutions of B$_z$ (vertical) and B$_\varphi$ (azimuthal) are shown in two lower panels in Figure \ref{fig2}. At the time of the flare, there is a clear indication of increase
in the amplitude of the vertical field and the decrease in the amplitude of the azimuthal component. The
radial $B_R$ component (not shown) did not change significantly as the result of this flare. We
interpret these changes as an indication that after the flare the magnetic field in penumbra became
less twisted and more vertical,which is in general agreement with the expectations of the twist-removal
model.
 
\section{Flare-related magnetic field changes: X2.2 flare on 15 February 2011}  
\label{sec:x2flare}

An X-ray X2.2 flare occurred on 15 February 2011 in NOAA active region (AR) 11158. For this flaring region, both LOS and vector magnetograms were taken by SDO/HMI, and we use these observations to compare the changes in magnetic field twist derived from LOS data with those of true-vector field observations.
HMI 12-minute LOS and vector magnetograms in the form of Active Region Patches (HARPs), maps of
vertical and two horizontal field components \{$B_r$, $B_\theta$, $B_{\phi}$\} with pixel size 0.03
degrees in heliographic coordinates derived by cylindrical equal-area (CEA) projection
\citep[see] [] {Hoeksema14}, have been used for this flare analysis. Here, $B_\theta$ and $B_{\phi}$ are the field components in the meridional and zonal direction, respectively. 

We transform the transverse true-vector field \{$B_\theta$, $B_{\phi}$\} to the radial and azimuthal
field components \{$B_R^\mathsf{v}$, $B_{\varphi}^\mathsf{v}$\} in cylindrical geometry as
\citep[see] [] {Venkatakrishnan09}:

\begin{eqnarray}
B_R^\mathsf{v} & = & \frac{1}{r} (xB_\phi + yB_\theta^\prime) \nonumber\\
B_\varphi^\mathsf{v} & = & \frac{1}{r} (-yB_\phi + xB_\theta^\prime) \nonumber
\end{eqnarray}

where $x$ and $y$ equal zero in the sunspot center, and $B_\theta^\prime=-B_\theta$ \citep[see] [for
details] {Sun13}. The vertical field component $B_z = B_r$. Here, $B_r$ refers to \cite{Venkatakrishnan09} notation for vertical component, which is $B_z$ in our notation. 

We obtain azimuthal averages of the true-vector \{$B_z^\mathsf{v}$,$B_R^\mathsf{v}$,
$B_{\varphi}^\mathsf{v}$\} and pseudo-vector \{$B_z$, $B_R$, $B_{\varphi}$\} magnetic field
components using Eq. \ref{eq2} as a function of distance from sunspot center, $r$.

\begin{figure}[h]
\begin{center}
\plotone{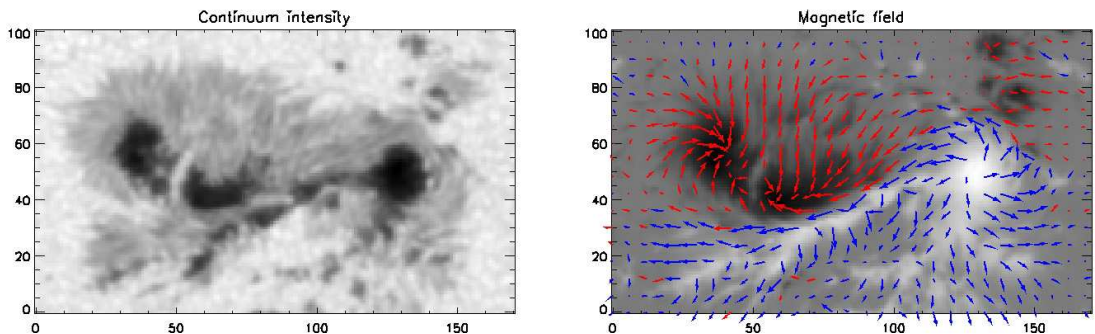}
\caption{Continuum intensity (left) and vertical magnetic field in the active region NOAA 11158 with the horizontal field shown with arrows (right).}
\label{fig3}
\end{center}
\end{figure}

The active region NOAA 11158 had a complex structure with four sunspots. Two of them were involved in
the flare. This part of the active region with two sunspots is shown in Figure \ref{fig3}. The
shapes of the sunspots can be partly considered as round with complex inclusions in their penumbras.
Based on the distribution of horizontal fields (Figure \ref{fig3} right), the sunspot of positive
polarity (white halftones) appears reasonably symmetric, and thus, is considered to be suitable for the application of the cylindrical symmetry approach. The sunspot of negative polarity (left side of panel) has an
elongated shape, and its umbra is divided by a light bridge. Still, based on the orientation of the
horizontal field, the portion of the sunspot to the left of the light bridge appears quite symmetric,
and thus, the assumption of cylindrical symmetry was applied to this sunspot too.

\begin{figure}[h]
\begin{center}
\plotone{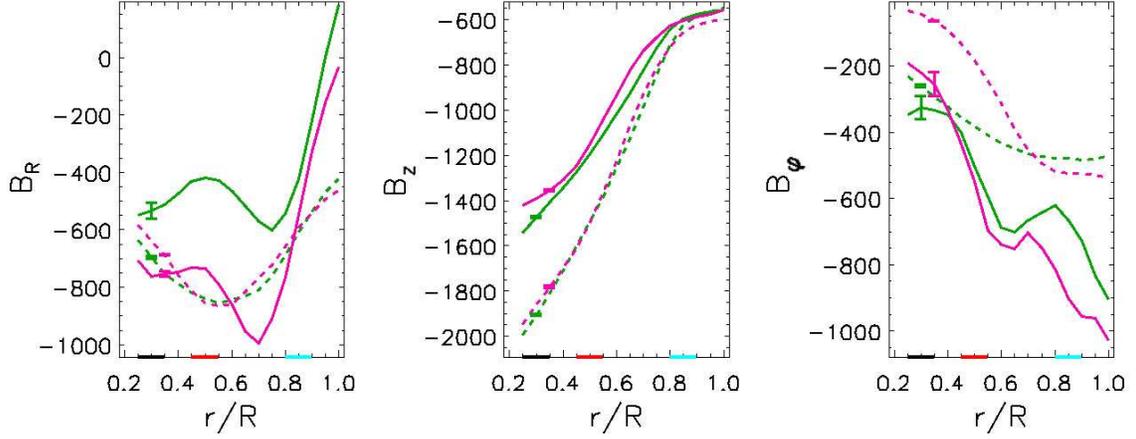}
\caption{Pseudo-vector \{$B_z$, $B_R$, $B_{\varphi}$\} (solid line) and true-vector
\{$B_z^\mathsf{v}$,$B_R^\mathsf{v}$, $B_{\varphi}^\mathsf{v}$\} (dashed line) magnetic field
components as a function of distance from the sunspot center in fractions of its radius, obtained
for the negative polarity sunspot located at the flare site in the active region NOAA11158 on
February 15, 2011 right before (green) and after (magenta) the flare. Error bars represent typical
uncertainties of the measurements. Black, red, and cyan dashes along $x$-axes indicate umbra,
umbra-penumbra boundary and penumbra regions, respectively, for temporal profiles in Figure 5.}
\label{fig:profB}
\end{center}
\end{figure}

Profiles of the pseudo-vector \{$B_z$, $B_R$, $B_{\varphi}$\} along with azimuthally averaged true-vector
\{$B_z^\mathsf{v}$,$B_R^\mathsf{v}$, $B_{\varphi}^\mathsf{v}$\} field components in the negative
polarity sunspot at pre-flare and post-flare times are shown in Figure \ref{fig:profB}. Average
profiles of the vertical B$_z$ component (Figure \ref{fig:profB}, middle) derived from Eq. \ref{eq2}
(pseudo-vector, solid lines) and by azimuthal averaging of true vector (dashed lines) show similar
changes with distance from the center of the sunspot (r/R = 0). Still, the B$_z$ from the pseudo-vector shows much weaker field strengths in umbral area (1400-1600 Gauss vs. 2000 Gauss), and they decrease less steeply with the radial distance from the sunspot center. Both approaches give similar average field strengths at the outer boundary of the sunspot (about 600 Gauss). The radial B$_r$ and azimuthal (tangential) B$_\varphi$ components show more differences between pseudo- and true-vector data. The true-vector component B$_R$ gradually increases from umbra to the umbra-penumbra boundary, and it decreases outward to the outer penumbra boundary (r/R =1). The pseudo-vector shows a slight decrease in the radial (with respect to radial direction away from sunspot center) component of the field, and it significantly increases at approximately middle of the penumbra. In the sunspot umbra, the azimuthal (tangential) component derived from assumption of cylindrical symmetry has a similar amplitude to the azimuthally averaged B$_\varphi$ from the true vector, but it increases much more steeply toward the outer penumbral sunspot boundary. Comparing the amplitudes of the vertical and
horizontal fields, on average, the true-vector field yields a less inclined vector (about 30-35 degrees
relative to the vertical direction at the umbra-penumbra boundary and 40-45 degrees in the middle of
the penumbra) as compared with the pseudo-vector (50-55 degrees in the middle of the penumbra). Thus, despite the overall agreement in sign of the three components of the vector field and similar behavior
(decrease-increase) in the radial direction from the sunspot center to its outer boundary, the profiles of
the three components look quite different between the two methods in their amplitudes and the steepness
of center-to-boundary variation. Similar differences could be noted in pre-/post-flare changes. The absolute and fractional (relative to true-vector) discrepancies in total flux of the sunspot between the pseudo- and true-vector method are about $2.66\times10^{17}$ Mx and 0.15, respectively. 

\begin{figure}[h]
\begin{center}
\plotone{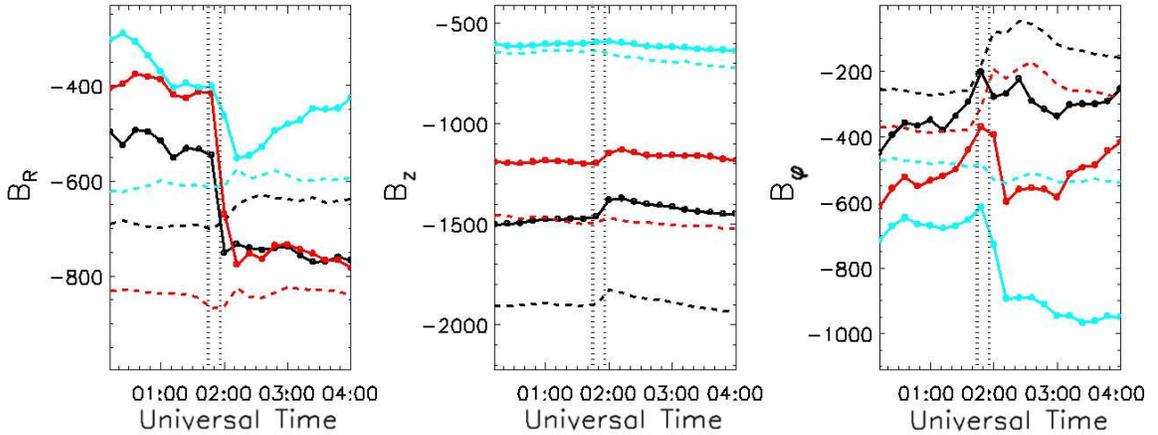}
\caption{Temporal profiles of the three vector-field components in the NOAA active region 11158 on
Feb 15, 2011 around time of the X2.2 flare. The results shown for the negative polarity sunspot
located at the flare site. Solid curves correspond to the pseudo-vector and dashed curves obtained
from true-vector observations in umbra (black), at the umbra-penumbra boundary (red), and in penumbra
(cyan). Vertical dotted lines indicate GOES flare start, peak, and end times.}
\label{fig:flarechange}
\end{center}
\end{figure}

The vertical component (Figure \ref{fig:flarechange}, middle panel) of the true-vector field (dashed lines)
shows some evidence of a slight decrease in the sunspot umbra as the result of the flare. The pseudo-vector data show a similar decrease, but, unlike the true-vector, such a decrease is observed in both umbra and penumbra. The largest difference between the two approaches can be seen in the B$_R$ and B$_\varphi$ components (left and right panels). The radial component in the pseudo-vector (left panel) exhibits strong increase after the flare, while the azimuthally averaged true vector does not show any significant
change in B$_R$.

The azimuthal component of the true-vector field decreases in the sunspot umbra and increases in the outer
penumbra as a result of the flare. This is consistent with the field becoming more horizontal in the
penumbra. The azimuthal (tangential) component of the pseudo-vector field decreases in the sunspot umbra in agreement with the true-vector measurements. $B_{\varphi}$ from the pseudo-vector increases in the outer penumbra similarly to the true vector, although the latter does not show a strong stepwise change as the former does. Overall, the changes in the three components of the vector field derived by two methods
could be interpreted as an indication of reduction in twist in umbral fields although the evolution
of penumbral fields is less clear in that respect.

No clear change in global properties of the field of the positive polarity sunspot in relation to the flare was found in either the pseudo-vector or the true-vector magnetic fields. Absence of the field changes in both true- and pseudo-vector field components suggests that flare-related changes could be local in their nature, and when averaged over the entire sunspot, these changes may not always be obvious.

\section{Distribution of magnetic field in isolated round sunspots}
\label{sec:round}

\begin{figure}[h]
\begin{center}
\plotone{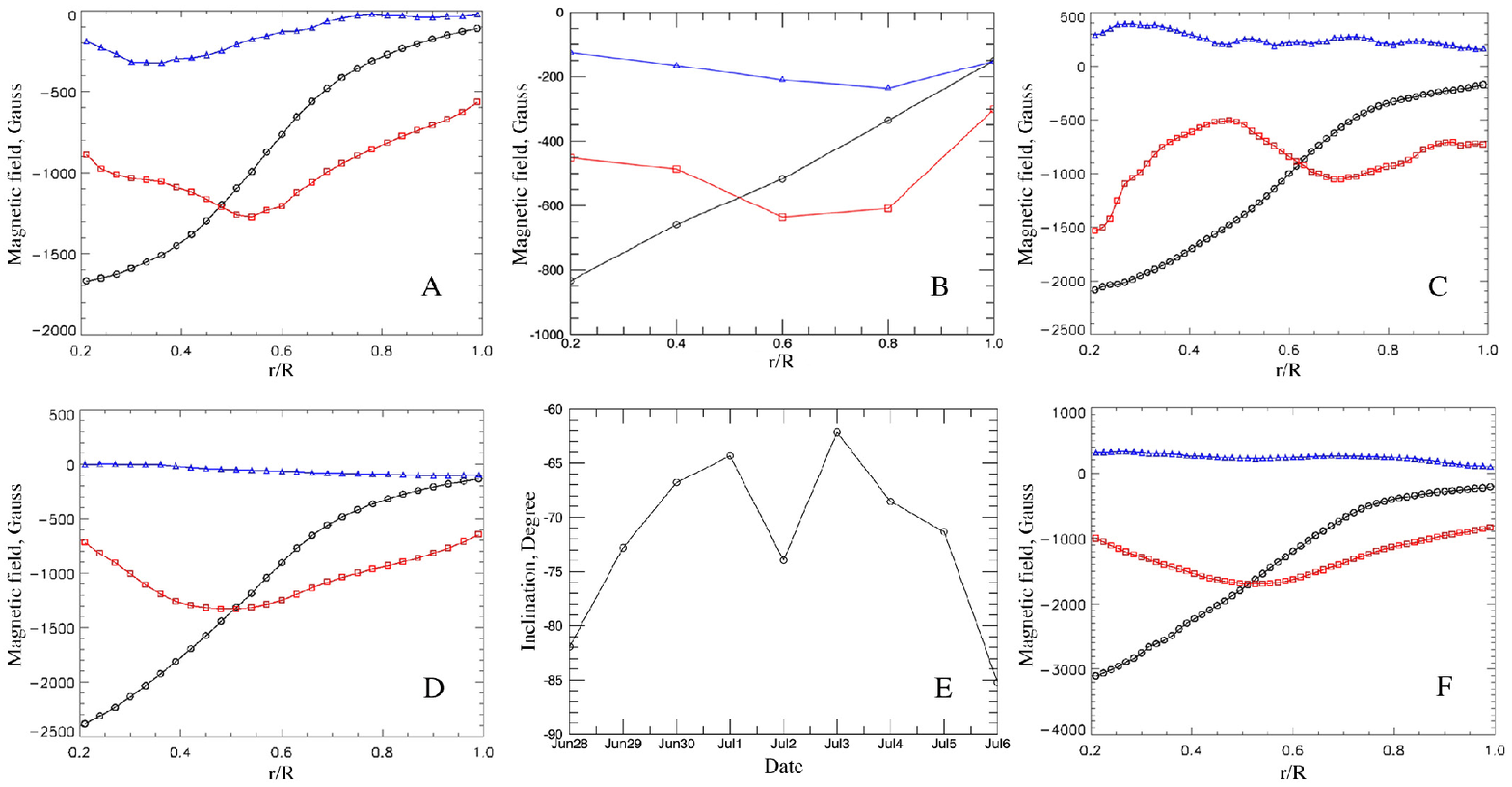}
\caption{Three components of magnetic field as a function of radial distance from sunspot center (R=0)
to sunspot outer penumbra boundary (R=1). Pseudo-vector magnetic field components $B_z$ (black
circles), $B_R$ (red squares), and $B_{\varphi}$ (blue triangles) in cylindrical symmetry obtained
from (A) HMI and (B) SOLIS/VSM photospheric longitudinal magnetic field in the NOAA11084 on July 1,
2010. (D) Azimuthally averaged magnetic field components as a function of distance from the sunspot
center in fractions of its radius obtained from HMI vector data. (C) and (F) Pseudo-vector and
true-vector magnetic field components, respectively, for the NOAA11899 on Nov 20, 2013. (E)
Inclination angle in NOAA11084 computed in the penumbra at R=0.75 of the sunspot's radius as the active region crosses the solar disk.}
\label{fig6}
\end{center}
\end{figure}

Examples of flare-related changes in magnetic fields shown in Sections \ref{sec:x6flare} and
\ref{sec:x2flare} indicate that, at least in some instances, the approach of computing the pseudo-vector
field could yield useful information about the evolution of average twist in flaring regions, but
the applicability of the method needs to be carefully evaluated. The discrepancies between the true- and pseudo-vector magnetic field in the flared active regions might arise due to the facts that the sunspots analyzed are not isolated, and they have asymmetries in their penumbras. For further validation of the cylindrical symmetry technique, we have selected 16 nearly round, isolated sunspots, listed in Table 1 (the X6.5 and X2.2 flare events are included in the Table as well). Fourteen of these sunspots were
analyzed using magnetograms taken for about one-hour time spans, usually around the time when the sunspot
was close to the solar central meridian. Two sunspots, a simple, round, uniformly twisted sunspot in
active region NOAA 11084 and another round sunspot in NOAA active region 11899, were studied over the period of a few days as they crossed the solar disk within $\sim 45^\circ$ heliographic longitude from
the solar disk center. The Pearson linear correlation coefficients for the azimuthal component of the pseudo- and true-vector magnetic fields for the 16 sunspots and their $P$-values, the probability that the observed correlation occurs by chance, computed using the t-test, are also shown in Table 1. Pseudo-vector \{$B_z$, $B_R$, $B_{\varphi}$\} along with azimuthally averaged
true-vector \{$B_z^\mathsf{v}$,$B_R^\mathsf{v}$, $B_{\varphi}^\mathsf{v}$\} field components for the
NOAA 11084 on July 1, 2010 and NOAA11899 on November 20, 2013 are shown in Figure \ref{fig6}.

\begin{table}[h]
\begin{center}
\caption{List of analyzed sunspots. Last column shows Pearson correlation coefficients (the $P$-values given in parentheses) for the azimuthal field component of the pseudo- and true-vector magnetic fields in the sixteen nearly round, isolated sunspots.}
\begin{tabular}{lcccl}
\tableline\tableline
Date                &  Location  &  NOAA     &  Flare    	     &	 Correlation Coefficient                                    \\
                        &                 &  Number  &   Event            &	 $B_{\varphi}$ vs. $B_{\varphi}^\mathsf{v}$\  \\
\tableline
2006 Dec  6     & S05E64   &  10930     &  X6.5              &                                                                         \\
2010 Jun 28 -  &  S19E45  &  11084      &  B1.0 (Jul 1)   &  -0.92 ($3.55\times10^{-15}$)	                      \\
       - Jul    6    &  S19W50  &                 &                        &                                                                         \\
2010 Aug  4     &  N12W07  & 11092      &                        &  0.61 ($5.24\times10^{-5}$)		              \\
2010 Aug 31    &  N12W05  & 11101      &                        &  -0.06 (0.76)		                                       \\
2010 Oct 21     &  S29W01   & 11115     &  		      &	  0.81 ($6.90\times10^{-9}$)                            \\
2010 Dec 10    &  N31W21  & 11133      &   		      &	  0.48 ($2.29\times10^{-3}$)                            \\
2011 Jan 21    &  N24E05  & 11147        & 			      &	  0.62 ($1.01\times10^{-3}$)                            \\
2011 Feb 15    &  S20W10  & 11158       &  X2.2 	      &	                                                                        \\
2011 Apr 19    &  N17E01   & 11193        &			      &	  0.79 ($4.81\times10^{-8}$)                            \\
2011 Apr 24    &  S16W01  & 11195        &			      &  -0.36 ($5.34\times10^{-2}$) 	                      \\
2011 Oct 10     &  N23W01  & 11312       & 		      &	  -0.22 (0.24)                                                    \\
2011 Nov 10    &  S08W01  & 11340       & 		      &	  0.09 (0.69)                                                      \\
2012 Mar 29     &  S23E07   & 11445      &  		      &	  0.88 ($1.48\times10^{-12}$)                           \\
2012 Sep 30    &  S10W01  & 11579       &			      &	  0.91 ($8.66\times10^{-15}$)                           \\
2012 Oct  18    &  N07E05  & 11591        & 		      &	  -0.23 (0.24)                                                     \\
2013 Nov 15 -  &  N07E37  & 11899        &		      &	  0.73 ($4.64\times10^{-11}$)                            \\
       - Nov 20    &  N05W29  &                  &		      &	                                                                         \\
2014 Feb 13    &  S15E11  & 11976         &		      &	  0.15 (0.41)                                                      \\
2016 Apr 26    &  S02W03  & 12533        &		      &	  0.57 ($2.85\times10^{-3}$)                             \\
\tableline
\end{tabular}
\end{center}
\end{table}

$B_R$ and $B_z$ components of pseudo-vector and azimuthally averaged true-vector field profiles seem
to be in a good agreement with each other. In most of the sunspot (R $<$ 0.8) from NOAA 11084, the amplitude of the tangential (azimuthal) component of the magnetic field derived under the assumption of cylindrical symmetry ($B_{\varphi}$) is larger than the azimuthally averaged azimuthal component of the true-vector field ($B_{\varphi}^\mathsf{v}$). The true-azimuthal field $B_{\varphi}^\mathsf{v}$ is close to zero near the sunspot center and then starts rising slowly, which is opposite to the trend of the pseudo-azimuthal field $B_{\varphi}$ in the sunspot umbra and inner penumbra, resulting in a negative correlation between them (see Table 1). There was a small (B1.0) X-ray flare in NOAA11084 on July 1 at 20:54 UT. Changes in inclination angle in NOAA11084 as it passed the solar disk, shown in Figure \ref{fig6}, seem to suggest that the magnetic field in the sunspot penumbra became slightly more vertical after the flare (inclination angle decreases), and then becomes more horizontal (inclination angle increases) again during the time span of one (one and half) day, which is in qualitative agreement with the twist-removal model, when helicity (twist) removed from a flaring/CME source active region can be replenished from below the photosphere within one or two days \citep[e.g.,][]{Pevtsov08}.   

For active region NOAA 11899 (Figure \ref{fig6}, panels (C) and (F)) the sign and the general trend of azimuthal $B_\varphi$ components of both the true- and pseudo-vector fields are in good qualitative agreement with each other. B$_R$ derived from cylindrical symmetry shows a more complicated variations with radial distance from sunspot center, as compared with $B_R^\mathsf{v}$ (Figure \ref{fig6}, panels (C) and (F)).

\begin{figure}[h]
\begin{center}
\plotone{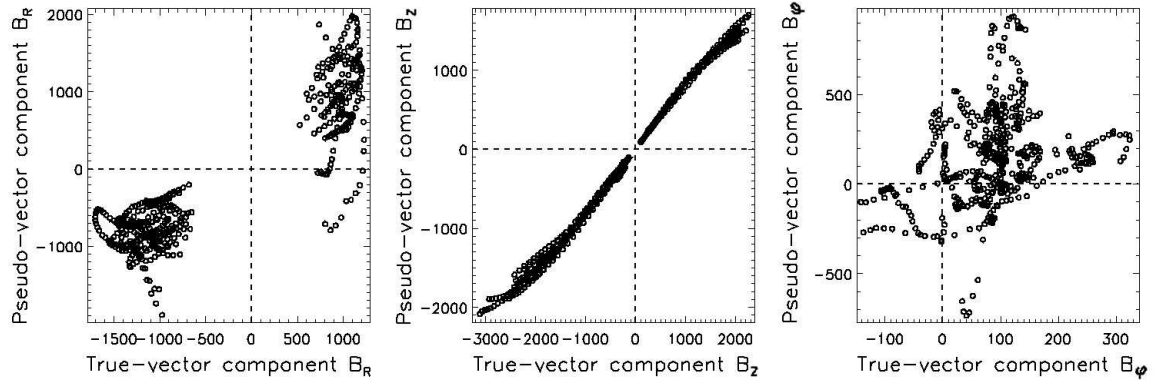}
\caption{Scatter plot of the pseudo-vector as a function of the true-vector magnetic field values, computed along the radial direction in the 16 round, isolated sunspots listed in Table 1.}
\label{fig7}
\end{center}
\end{figure}

Analysis of other 14 round sunspots indicates that the cylindrical symmetry approach does, in
general, provide a correct sign of the pseudo-vector field components, but the trend of the pseudo-vector components as a function of radial direction from sunspot center can be different from that derived from the true-vector field. Also, the amplitude of the pseudo-vector field variations along the sunspot radius is often much larger than the variations in the true-vector field. Scatter plots of the true- and pseudo-vector field values computed along the radial direction in the 16 isolated sunspots in Figure \ref{fig7} summarize the results. The Pearson linear correlation coefficients for the $B_R$, $B_z$, and $B_{\varphi}$ components of the pseudo- and true-vector magnetic fields are 0.90, 0.99, and 0.31 and their $P$-values are $0.0$, $0.0$, and $1.8\times10^{-12}$, respectively, revealing a strong correlation for the $B_R$ and $B_z$ components and weak correlation for the $B_{\varphi}$ component, in general. The $B_{\varphi}$ component of the pseudo- and true-vector magnetic fields in only half of the 16 sunspots reveals a statistically significant positive moderate or strong correlation (see Table 1).

\section{Discussion} 
\label{sec:discussion}

In this work, we analyzed magnetic field changes associated with two flares, the X6.5 flare on 6
December 2006 and X2.2 flare on 15 February 2011. In particular, we used the data to search for a
signature of removal of twist as the result of flares. We applied the azimuthal symmetry approach to
investigate the changes in twist and inclination of magnetic field in two sunspots that hosted major X-ray flares (X6.5 flare in AR NOAA 11084 and X2.2 flare in AR NOAA 11158). For the latter active region, we
compared the results with the azimuthally averaged true-vector field. For both active regions, we found the changes in orientation of magnetic field consistent with the twist-removal scenario.

However, our study also raises some questions about the applicability of the cylindrical symmetry approach.
While the pseudo-vector field revealed some of the abrupt changes in the magnetic field associated with X-class flares, the amplitude and/or the sign of these changes does not always clearly agree with
changes derived from true-vector field data. The shapes of the sunspots can only partly be considered as round. The cylindrical (azimuthal) symmetry approach could be more readily justified
in isolated round sunspots, but even for such sunspots, the derived pseudo-vector components may
disagree with azimuthally averaged true-vector field components. Absence of clear flare-related
field changes in both true- and pseudo-vector field components for one of the sunspots suggests that
flare-related changes could be local in their nature, and when averaged over the entire sunspot,
these changes may become less apparent.

\begin{figure}[h]
\begin{center}
\plotone{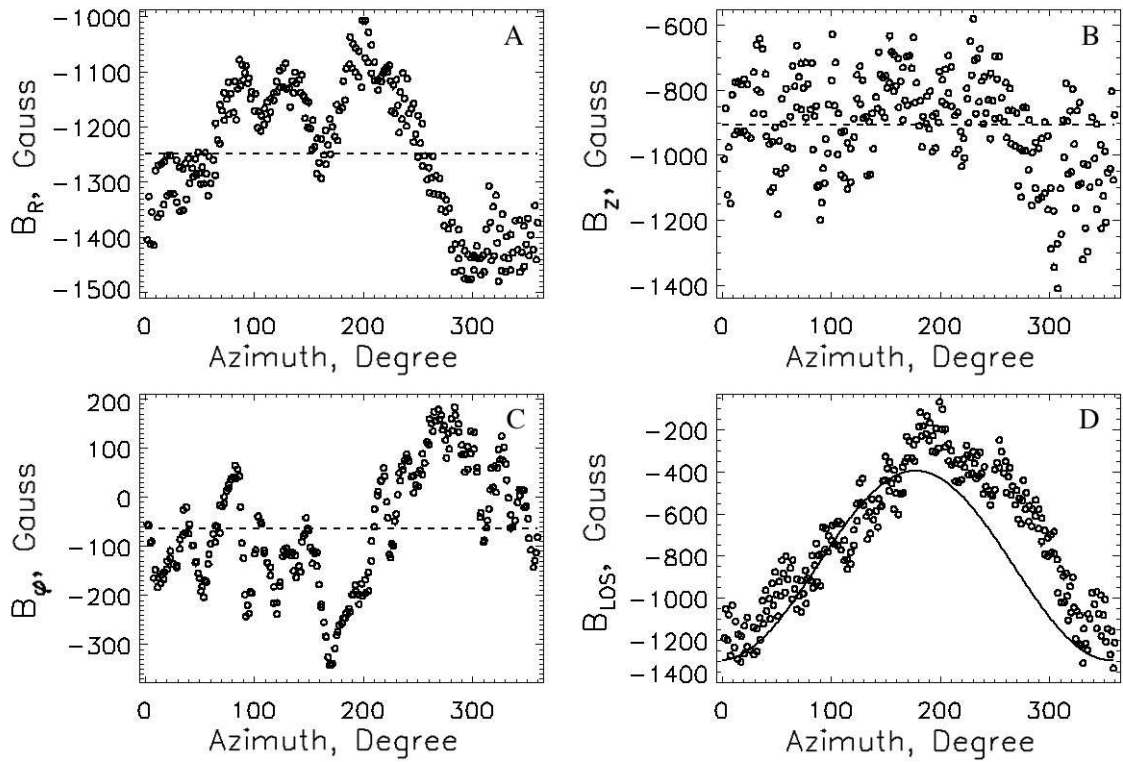}
\caption{Variation of true-vector \{$B_z$, $B_R$, $B_{\varphi}$\} and LOS magnetic field components
as a function of azimuth at R=0.6 of the sunspot's radius for the sunspot in the active region
NOAA11084 on July 1, 2010. Horizontal dashed lines in panels (A)--(C) indicate the mean value for each
component. The solid curve in panel (D) shows the expected variation in B$_{LOS}$ computed using mean $B_z$, $B_R$, and $B_{\varphi}$ and Equation \ref{eq1}.}
\label{fig8}
\end{center}
\end{figure}

Finally, we have investigated how well the azimuthal (cylindrical) symmetry approach works for  isolated round sunspots. We found that the results from this approach agree with true vector for selected sunspots, but even for symmetric sunspots the distribution of twist may be complex. Figure \ref{fig8}
(panel (D)) shows the distribution of the longitudinal (LOS) component of the magnetic field as a function of
azimuthal angle at a fixed distance from the center of a sunspot. The variation of B$_{LOS}$ with
azimuth does change in a manner similar to the expected behavior (namely, as
if at different azimuths, B$_{LOS}$ represents the LOS projection of the same vector field).
However, this variation does not agree exactly with changes one would derive using Eq. \ref{eq1} 
under the assumption that B$_z$, B$_r$, and B$_\varphi$ components of that vector are constant (solid line in Figure \ref{fig8}d). Figure \ref{fig8} (panels (A)--(C)) shows the variation of each component of a true vector as a function of azimuth. The horizontal dashed line in each panel shows the mean value of this vector field component. It is clear that the azimuthal distribution of each of the three vector field
components is not constant for a fixed radial distance from sunspot center. Out of three components,
B$_z$ (vertical) is approximately constant for range of azimuths between 0 and 260 degrees. Between
260 and 360 degrees, the mean value is quite different. Two other components show even less
constancy with azimuthal angle inside the sunspot. For this sunspot, one could argue that B$_r$ could
still be considered ``constant'' within $\pm$ 15-20\%. However, the azimuthal (B$_\varphi$)
component even has opposite signs in different parts of the penumbra. These shortcomings of cylindrical (azimuthal) symmetry need to be taken into consideration in all future attempts to employ this type of model.

\newpage
\section{Acknowledgments}
The HMI data are provided by NASA/SDO and the HMI Science Team. GONG and SOLIS data obtained by the NSO Integrated Synoptic Program (NISP), managed by the National Solar Observatory, which is operated by the Association of Universities for Research in Astronomy (AURA), Inc. under a cooperative agreement with the National Science Foundation. Authors of this work are partially supported by NASA grant NNX14AE05G. A.A.P. acknowledges the financial support by the Academy of Finland to the ReSoLVE
Centre of Excellence (project No. 272157).

\bibliography{Burtseva_etal_bib}

\end{document}